\documentclass[american,aps,pra,reprint,superscriptaddress,nofootinbib, floatfix]{revtex4-1}
\usepackage[T1]{fontenc}
\usepackage[utf8]{inputenc}
\usepackage{color}
\usepackage{babel}
\usepackage[final]{graphicx} 
\usepackage{color}
\usepackage{tabularx}
\usepackage{braket}
\usepackage{multirow}
\usepackage{booktabs}
\usepackage{amsmath, mathtools}
\usepackage{amssymb}
\usepackage[textsize=small, obeyDraft]{todonotes}
\usepackage{ifdraft}
\usepackage[unicode=true,pdfusetitle,
 bookmarks=true,bookmarksnumbered=false,bookmarksopen=false,
 breaklinks=false,pdfborder={0 0 0},backref=false,colorlinks=true, draft=false]
 {hyperref}

\graphicspath{{images/}}
\usepackage{times}
\usepackage{braket}
\usepackage{enumerate}
\usepackage{colortbl}

\newcommand{\CZ}{\mathit{CZ}}

\begin{document}
\ifdraft{\setlength{\marginparwidth}{\columnwidth}}{}

\title{Distributing Graph States Over Arbitrary Quantum Networks}

\author{Clément Meignant}

\address{Laboratoire d'Informatique de Paris 6, CNRS, Sorbonne Universit\'e, 4 place Jussieu, 75005 Paris, France}

\author{Damian Markham}

\address{Laboratoire d'Informatique de Paris 6, CNRS, Sorbonne Universit\'e, 4 place Jussieu, 75005 Paris, France}

\author{Frédéric Grosshans}

\address{Laboratoire Aimé Cotton, CNRS, Univ. Paris-Sud, ENS Cachan, Univ. Paris-Saclay, 91405 Orsay Cedex}

\address{Laboratoire d'Informatique de Paris 6, CNRS, Sorbonne Universit\'e, 4 place Jussieu, 75005 Paris, France}

\begin{abstract}
    Multipartite entangled states are great resources for quantum networks. In this work we study the distribution, or routing, of entangled states over fixed, but arbitrary, physical networks. Our simplified model represents each use of a quantum channel as the sharing of a Bell pair; local operations and classical communications are considered to be free.
    We introduce two protocols to distribute respectively Greenberger--Horne--Zeilinger (GHZ) states and arbitrary graph states over arbitrary quantum networks. The GHZ states distribution protocol takes a single step and is optimal in terms of the number of Bell pairs used;
    the graph state distribution protocol uses at most twice as many Bell pairs and steps than the optimal routing protocol for the worst case scenario.  
\end{abstract}
\maketitle

\section{Introduction and setting}

    Classical networks are actively present in many areas of day-to-day life. 
    Whether on a global scale, with world-wide communication or, on a much smaller scale, on multi-processor devices distributing computation over several processors \cite{sunderam1990pvm}, these networks distribute, share and use information. 
    The study of quantum networks is a recent active field of quantum information with promising applications. 
    These range from secure communication, clock synchronization, exponential gains in communication complexity, distributed sensing to delegated computation in the cloud, distributed computation and more.
    Distributing quantum states over all kinds of quantum networks is a necessary step to implement most of these applications and will consume quantum resources which may be difficult to replenish. 
    It is thus necessary to find ways of distributing quantum states while minimizing the cost. 
    Until recently, most of the work published about quantum networks and entanglement routing concerned point-to-point communications and bipartite setting \cite{schoute2016shortcuts, acin2007entanglement, perseguers2008entanglement,  pirandola2016capacities}, with a few recent exceptions \cite{epping2016large, cuquet2012growth, matsuzaki2010probabilistic, pirker2018modular, HahnPappaEisert2018, pirker2018quantum}.

As a simplified model, we represent here a quantum network by
a graph, an example of which is depicted in Figure~\ref{fig:quantum_network}. 
Nodes (the dotted circles) represent physical locations in the network. 
Within these nodes, local computations (restricted to Clifford operations \cite{van2004graphical} in our protocols) are considered free.
Quantum channels between nodes are represented as shared Bell pairs%
---pictured as solid vertices with edges between them.
We consider classical communications to be free, hence the Bell pairs can be considered as single uses of the quantum channel.
Note that each node may contain several qubits, but each edge corresponds to a single
Bell pair.
These Bell pairs can be replenished at each step in our protocols, but only along the original edges of the network (representing the physical quantum channels).
Our goal is to distribute entangled states across this network in a way that is most efficient in terms of the number of Bell pairs consumed and the number of steps taken.

\begin{figure}
	\caption{A quantum network, each vertex represents a qubit and each edge 
		represents a Bell pair; each dotted circle represents a node of the network.
		Note that a node can hold several qubits.}
	\label{fig:quantum_network}
	\centering
	\includegraphics[width=0.70\columnwidth]{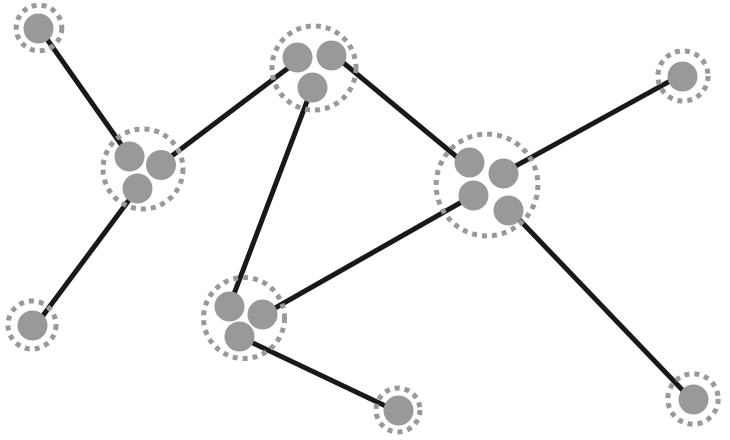}
\end{figure}
The use of such a network to create a maximally entangled bipartite state between two distant nodes for point-to-point communications has been extensively 
studied in the literature \cite{acin2007entanglement, perseguers2008entanglement, schoute2016shortcuts, pirandola2016capacities}.
This network can also be used to share multipartite entangled states, either by simultaneously sharing several entangled pairs between different sets
of clients \cite{Pant2019Routing, khatri2019practical}, or by sharing a genuinely multipartite state---%
a useful resource for quantum communication \cite{agrawal2006perfect, zheng2006splitting}, 
computing \cite{schlingemann2001quantum, raussendorf2003measurement, briegel2009measurement} or 
metrology \cite{chuang2000quantum} protocols. 

In this work, we will study the distribution of graph states---a large class 
containing many useful multipartite entangled states \cite{agrawal2006perfect,GraphsLong}%
---over networks of arbitrary topology. 
To simplify the study, we ignore  the cost of classical communications and we also neglect
 the processing time of the local quantum processors and the cost of memory. 
We will also assume the distribution of Bell pairs to be perfect and to occur at perfectly synchronized times, and the node local computations to be similarly perfect.
Note that operations between several qubits of the same node are considered here to be local; a wider range of graph states transformations are thus available compared to other work tackling the manipulation of graph states \cite{van2004graphical, HahnPappaEisert2018}.

The distribution of multipartite entangled states over quantum networks has also been studied in 
\cite{epping2016large, cuquet2012growth, matsuzaki2010probabilistic, pirker2018modular, HahnPappaEisert2018, pirker2018quantum}. 
In \cite{epping2016large}, the authors investigate the creation of a graph state presenting the shape of the network in the presence of noise. References \cite{cuquet2012growth, matsuzaki2010probabilistic} present decomposition of graph states into various building blocks that can be purified and merged to construct graph states over a network.
Reference \cite{HahnPappaEisert2018} studies the possible transformations of an already shared graph-state, with a single-qubit per location. 
Reference \cite{pirker2018modular} presents a modular architecture to fulfill graph states creation requests. 
To our knowledge, with the exception of \cite{pirker2018quantum},
no work has been published about the complete process of sharing entangled states from scratch as outlined here.
During the redaction of this manuscript we noticed the publication of independent work of Pirker and Dür \cite{pirker2018quantum}, which includes a protocol very similar to ours.
The modeling of the network in both works is different, as well as the optimized metrics.
They describe a hierarchical network stack and use it to provide robustness against router or sub-network failures, which we do not study.
However,
they do not have a cost metric, be it in entanglement use or time, and do not provide a statement of optimality as we do here.
 
 The article is organized as follows. We begin in section \ref{sec:statesandtools} by giving background on graph states and the graphical tools we use.
 Then, in section \ref{sec:Distribution}, we propose several protocols to distribute multipartite states  over arbitrary quantum networks using only basic operations of graph states, 
 starting in subsection \ref{sec:GHZ} with a minimal protocol distributing 
 Greenberger--Horne--Zeilinger (GHZ) states \cite{GHZ}. 
 In subsection \ref{sec:ArbGraph}, we then show a method to distribute arbitrary graph states by distributing several GHZ states over the network.
Then, in subsection \ref{sec:optimality}, we analyze the optimality of these protocols.
In section \ref{sec:imperfections}, we briefly discuss the different imperfections
overlooked in our idealized model and suggest ways to cope with them.
 We close in section \ref{sec:conclusion} with discussions.

\section{Graph States and Graphical Tools}
\label{sec:statesandtools}
The graph state $\ket{G}$ associated to the simple graph $G=(V,E)$, with vertices $a \in V$, and edges $a,b \in E$, is
\begin{align}
    \ket{G}&:=\prod_{\mathclap{(a,b)\in E}}\CZ_{a,b}\ket{+}^{V},&
    \text{where } \ket{+}^V&:=\bigotimes_{a\in V}\ket{+}^a
\end{align}
 is the tensor product of all qubits of $V$ in the state 
$\ket{+}^a:=(\ket{0}^a+\ket{1}^a)/\sqrt{2}$,
 and $\CZ_{a,b}$ is the controlled-$Z$ operation between qubits $a$ and $b$. 
That is, for a graph $G$, each vertex represents a qubit, and each edge a $\CZ$ entangling operation. 
Note that the quantum network represented in Figure \ref{fig:quantum_network} can be understood as a collection product of bipartite graph states, each of the pair represented being a Bell pair.

\begin{figure}
    \caption{Example of the application of local complementation applied on vertex $3$. The physical operation associated is written as $U_3^{\tau}=e^{-i\frac{\pi}{4}X_3}\otimes e^{i\frac{\pi}{4}Z_1} \otimes e^{i\frac{\pi}{4}Z_4} \otimes e^{i\frac{\pi}{4}Z_5}$. The edge between $1$ and $5$ is removed and two edges are created between both the pairs $(1,4)$ and $(4,5)$}
    \label{fig:quantum_complementation}
    \centering
    \includegraphics[width=.7\columnwidth]{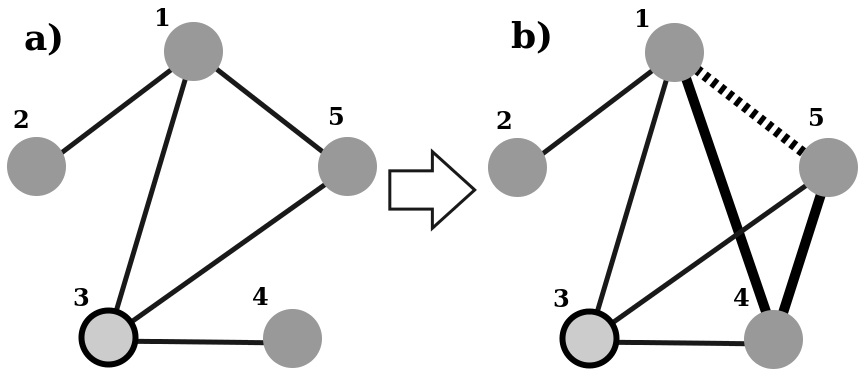}
\end{figure}

The work presented in this article is based on graphical representation of operations and measurements. 
Indeed, several physical operations on graph states $\ket{G}$ can be represented as graph operations on $G$  (up to local corrections that we will neglect here and in the following). In particular, we will use three elementary graph operations as building blocks for our protocols \cite{van2004graphical, MBQCTutorial}:
\begin{enumerate}[(i)]
    \item Vertex deletion. This operation removes one vertex and all the associated edges from the graph.
      Physically, it is implemented by the Pauli measurement of the relevant qubit in the $Z$ basis.
    \item Local complementation on a vertex. This graph operation inverts the 
      sub-graph induced by the neighborhood $N_a$ of the concerned vertex $a$%
      ---the set of vertices adjacent to $a$ (see Figure \ref{fig:quantum_complementation}). 
      It is implemented by applying the relevant operation to the qubits of $a\cup N_a$,
      described by the quantum operator 
      $U_a^{\tau}:=e^{-i\frac{\pi}{4}X_a}\bigotimes_{b\in N_a}e^{i\frac{\pi}{4}Z_{b}}$ 
      acting on $\ket{G}$. 
    \item Edge addition (deletion). By applying a controlled-$Z$ operation 
      between two qubits belonging to the same node, we create (delete) an edge 
      between two non-adjacent (adjacent) vertices.
\end{enumerate}
Another useful, if non-elementary, operation, is the measurement of a qubit in the $Y$ basis,
which corresponds graphically to a local complementation followed by the removal 
of the measured vertex. To see this, we note that the local complementation operations implement a basis change from  $Z$ to a $Y$ on the concerned vertex.

As an example, Figure \ref{fig:entanglement_swapping} shows how entanglement 
swapping along a line of repeaters \cite{schoute2016shortcuts, acin2007entanglement, perseguers2008entanglement} can be depicted graphically with the
above tools. The essential observation here is that a Bell measurement is 
equivalent to performing a $\CZ$-gate followed by two single qubit $Y$-measurements.

\begin{figure}
    \caption{Graphical representation of the quantum repeater protocol. 
    Starting from a repeater line, we apply $\CZ$ and two measurements in the $Y$ 
    basis at each repeater to obtain a Bell pair between the end nodes.}
        \centering
    \includegraphics[width=.7\columnwidth]{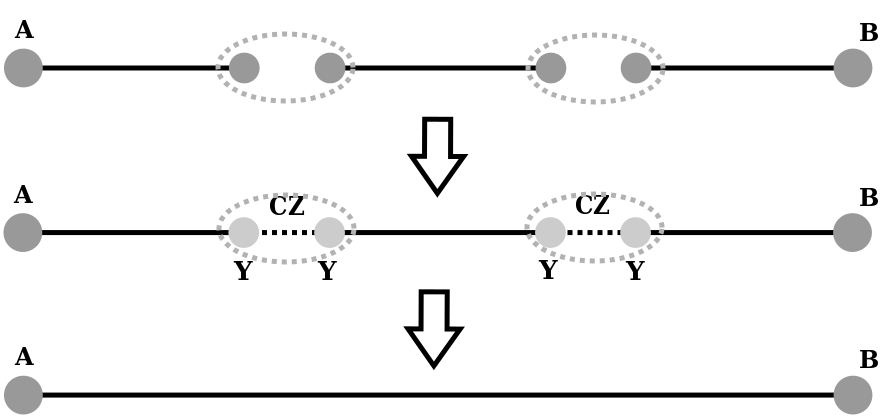}
    \label{fig:entanglement_swapping}
\end{figure}

\section{Graph state distribution protocols}\label{sec:Distribution}
We will now see protocols distributing graph states, starting with
a minimal protocol to distribute GHZ states over an arbitrary quantum network.

\subsection{GHZ State Distribution}\label{sec:GHZ}
GHZ states form a useful class of multipartite maximally entangled states. 
They are used in many multiparty applications of quantum information such as quantum secret sharing \cite{hillery1999quantum} or quantum metrology \cite{dur2014improved}. 
Thus, establishing a protocol to distribute them is an important step toward the implementation of multipartite protocols over a quantum network. 
A $N$-GHZ state is written $\ket{N\textrm{-GHZ}} = (\ket{0}^{\otimes N} + \ket{1}^{\otimes N})/\sqrt{2}$.
It is locally equivalent to 
$(\ket{0}\ket{+}^{\otimes N-1} + \ket{1}\ket{-}^{\otimes N-1})/\sqrt{2}$, 
the star graph with an edge linking the first qubit---the center---to each of the others.
The choice of which vertex is the center is arbitrary, and can be changed by 
local operations such as two successive local  complementations.
We will distribute the star graph on the network using our graphical rules.
Given an arbitrary set $W$ of the network's nodes, we will now see how to distribute a star graph among all the nodes of $W$. 
The amount of Bell pairs consumed is minimal and we can distribute it in one time-step.

This protocol relies on an operation we call \emph{star expansion}, which acts 
on the qubit $b$ of a graph-state --- $b$ will be the center of a star in our case --- 
and a node $A$ of the network
which contains a qubit $a_0\in A\cap N_b$ in the neighborhood of $b$.
Each of the other qubits $a_i\in A$, $i>0$ of $A$ constitutes a Bell pair with
a qubit $c_i$ in another node of the network.
The star expansion operation, detailed in Figure \ref{fig:entanglement_distrib}, 
uses the Bell pairs of the node $A$ to add the edges $(b,c_i)$ to the graph-state, 
as well as the edge $(b,a_0)$ iff $A \in W$.
    \begin{figure}
        \caption{The star expansion operation: 
            \textbf{a)} all qubits $a_i$, $i\ge0$ of $A$ are linked using $\CZ$ between all possible pairs;
            \textbf{b)} local complementation is applied to the qubit $a_0$ linked to $b$;
            \textbf{c)} if $A\notin W$, we remove this qubit and all edges within $A$ by 
               $Z$-measuring it;
            \textbf{e)} else, when $A\in W$, we keep $a_0$ and apply $\CZ$ gates to remove all edges within $A$;
            \textbf{d)},\textbf{f)} finally, a $Y$-measurement of all other qubits $a_i\in A$, $i>0$  creates
              the desired star graph.}
        \label{fig:entanglement_distrib}
        \centering
        \includegraphics[width=.95\columnwidth]{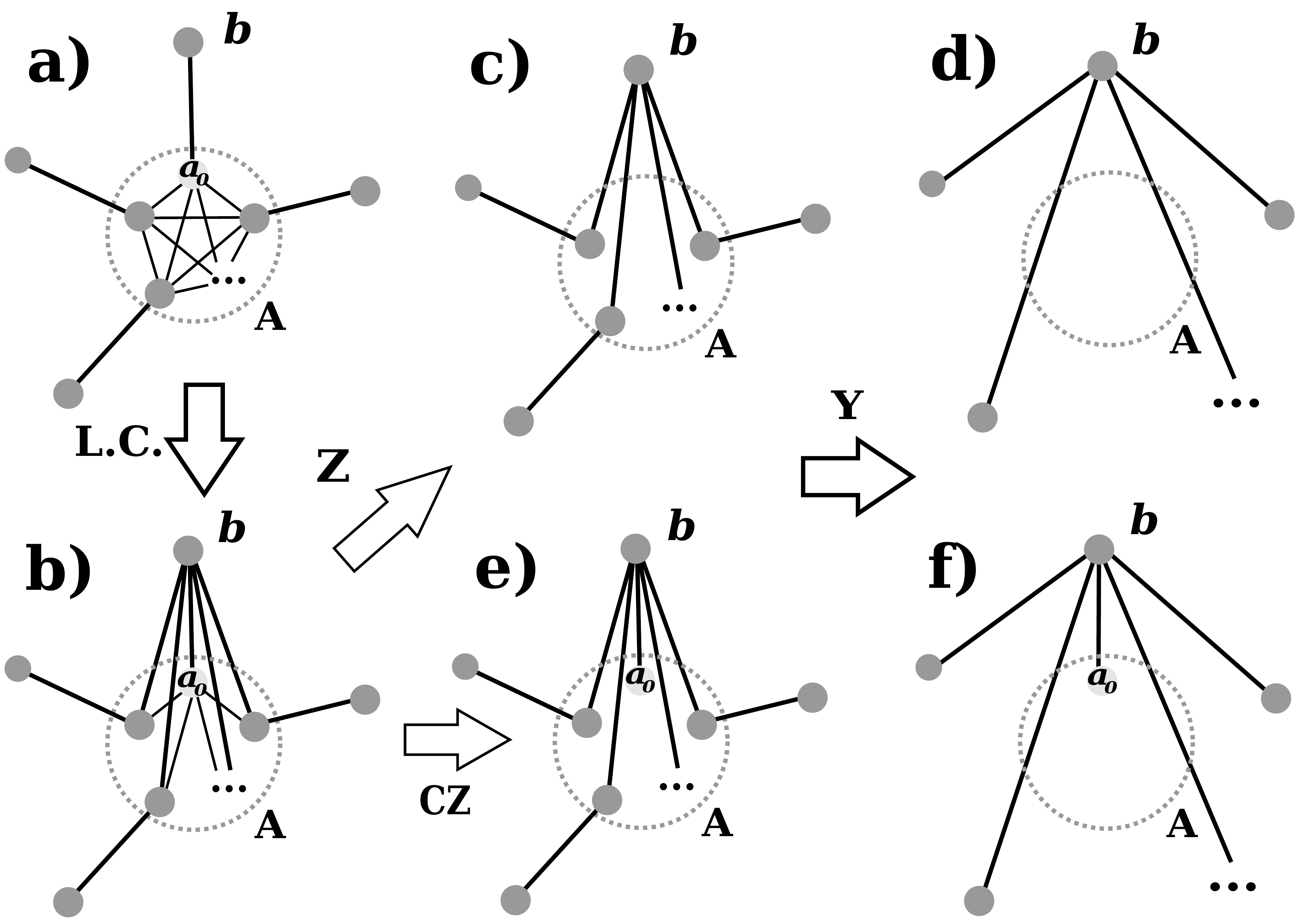}
    \end{figure}

The star expansion subprotocol defined above will help us to share the star graph 
state across the full set $W$.
The first step is to find a minimal tree covering all the nodes of $W$%
---i.e.\@ a subgraph connecting all the nodes in $W$ with the minimum number of edges. 
Such a problem is \textit{the Steiner tree problem}, well-known in classical graph theory.
Despite being NP-Hard \cite{hwang1992steiner}, 
the Steiner tree can be approximated in a polynomial time \cite{robins2005tighter, berman20091}. 
See Figure \ref{fig:steiner_tree} for an example Steiner tree defined on the network 
of Figure \ref{fig:quantum_network}.    
\begin{figure}
    \caption{Steiner tree example for a set of nodes $W$. The network vertices and edges are grey. The node of $W$ are represented by black-dotted circle. On the left is the original network (vertices and edges grey), and on the right, with black edges, is the associated Steiner tree for the set $W$.}
    \centering
    \includegraphics[width=.95\columnwidth]{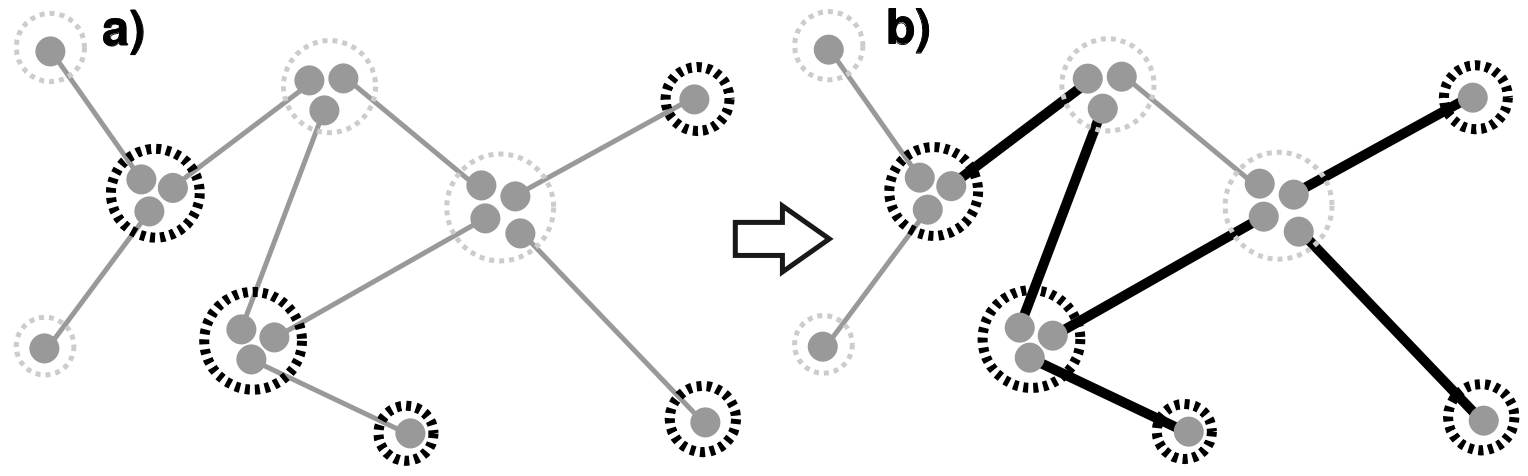}
    \label{fig:steiner_tree}
\end{figure}

Starting from any leaf $\ell$ of this tree, we can distribute the star graph 
by exploring the tree and applying star expansion with the exploration's current node (a non-leaf neighbor of $\ell$) as $A$, and $\ell$ as $b$. This process is depicted in Figure \ref{fig:GHZ_distrib}. 

\begin{figure}
    \caption{Distribution of a star graph state.  We distribute the star graph over the  nodes of $W$, represented by black-dotted circles. To explore the tree, we arbitrarily choose a non-leaf neighbour of $\ell$ and apply star expansion on that node $A$ (taking $\ell$ as Bob). We repeat this until we arrive at the desired star over $W$.}
    \centering
    \includegraphics[width=.95\columnwidth]{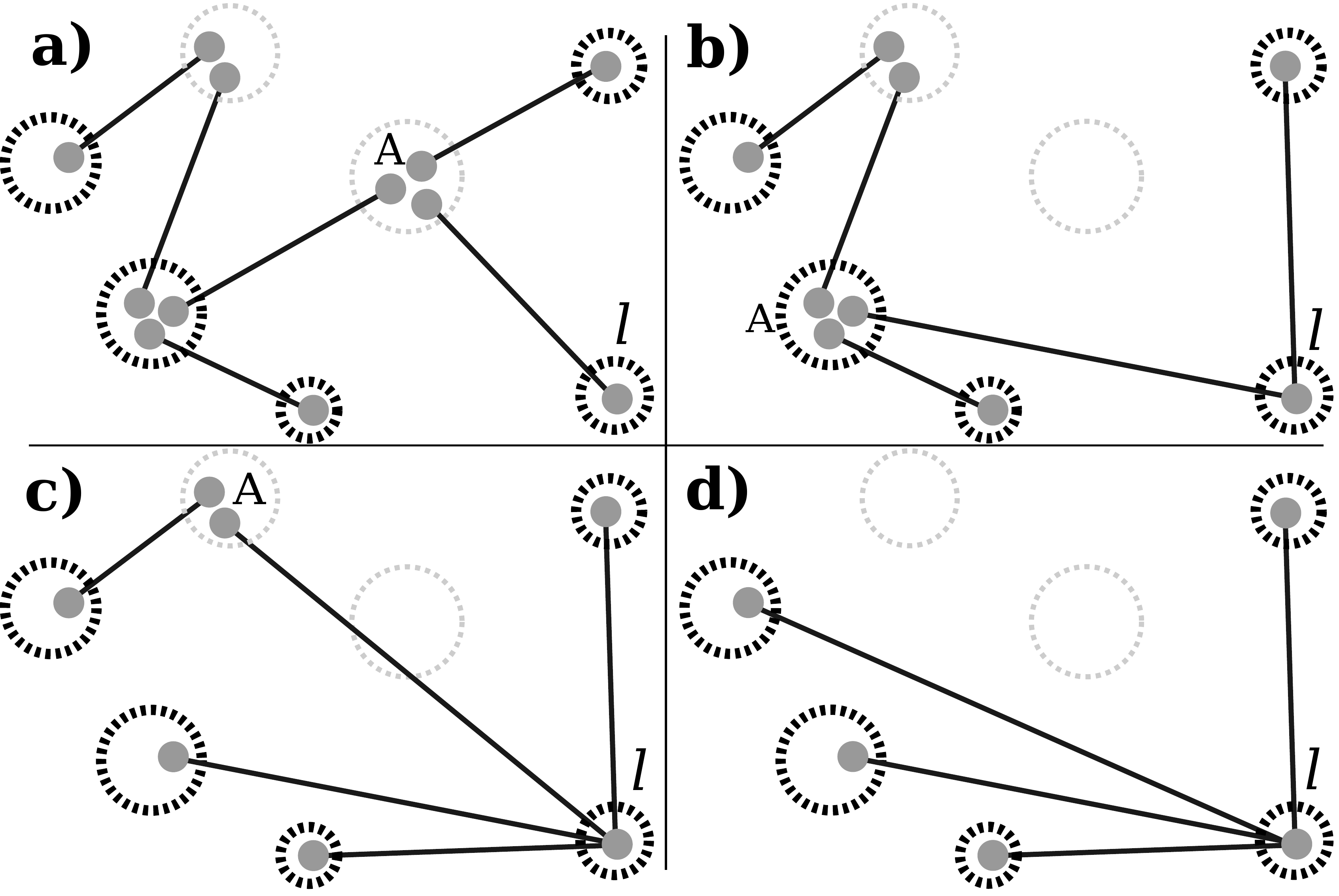}
        \label{fig:GHZ_distrib}
\end{figure}

The number of Bell pairs consumed in this process is equal to the number of edges in the tree, 
which by definition is (almost) the minimum possible number for the (approximate) Steiner tree. 
Since star expansion requires local operations to each node and the same Clifford operations at $\ell$ each time-step, these commute and can all be done in one step,
with a single step of correction afterwards \cite{browne09computational}.


\subsection{Arbitrary Graph State Distribution}\label{sec:ArbGraph}
We now show how to generalize the previous
approach to distribute an arbitrary graph state over a set $W$ of known nodes of the network. 
The procedure will be to distribute a specific resource graph state: the edge-decorated complete graph.
From this graph state, nodes can construct any graph state by measuring each edge-qubit in either the $Z$ basis or the $Y$ basis, as represented in Figure~\ref{fig:decorated_to_graph}.
This graph is already used in \cite{pirker2018modular} for a similar goal, with a protocol to 
distribute it in a different context. We present here a new approach to its distribution, 
adapted to our setting, as well as compute and optimize its cost in term of resources.

\begin{figure}
	\centering
	\caption{The edge-decorated graph can be projected into any graph state by measuring its edge-qubits. We can distribute it as a resource to generate arbitrary graph state.}
	\includegraphics[scale=0.20]{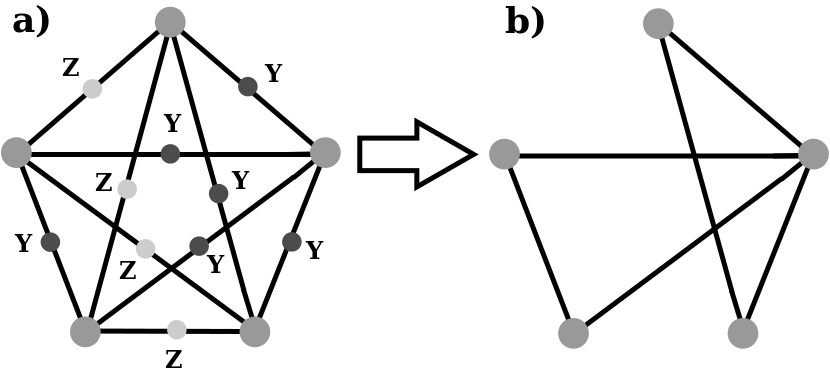}
	\label{fig:decorated_to_graph}
\end{figure}

The protocol to distribute the edge-decorated complete graph follows directly from multiple applications of the GHZ-state distribution. We consider the set $W$ of the $k$ participating nodes. The first step is solving the \textit{Steiner's Tree problem} on the network for the $k$ nodes. Then, we distribute a $k$-GHZ state starting from one arbitrary leaf $\ell_1$. We delete vertices from the tree in order to have the covering tree for the set $W\setminus\{\ell_1\}$ and we distribute from a leaf $\ell_2$ of this tree a $(k-1)$-GHZ state. This procedure iterates until the distribution of a final Bell pair between the two last nodes of $W$. 
As seen in Figure \ref{fig:decorated_distribution}, the resulting graph state is locally equivalent to the edge-decorated dotted graph.
\begin{figure}
    \centering
    \caption{Distribution of the edge-decorated graph of size 5 from a minimal tree (Steiner Tree) starting with \textbf{a}) at T=0 . At each step, a star graph is shared centred at $\ell_i$ (subsequently indicated in grey), then vertex $i$ is ignored in the following steps. Finally, in \textbf{f}) node local operations generate the desired edge-decorated graph state.}
    \includegraphics[width=\columnwidth]{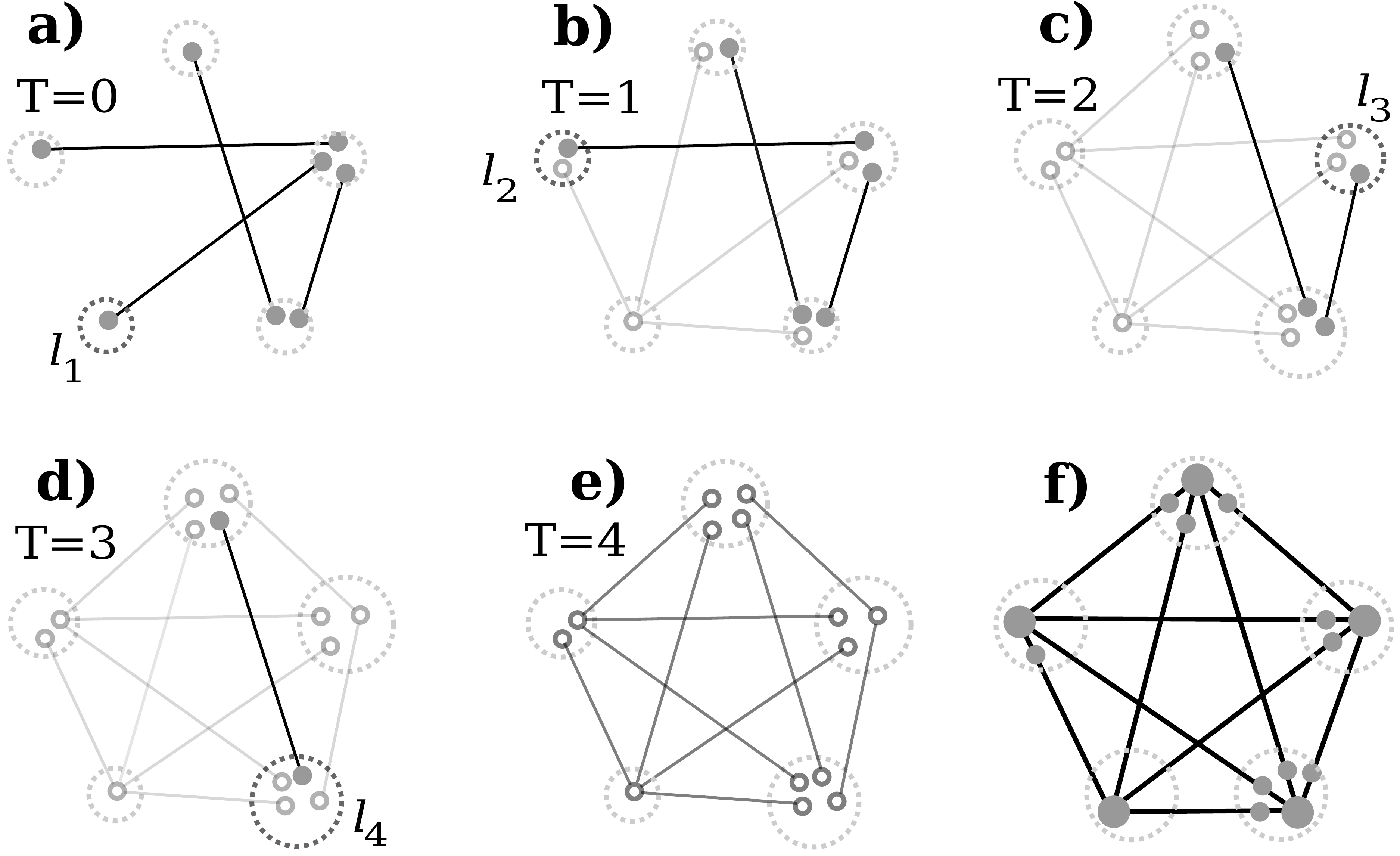}
    \label{fig:decorated_distribution}
\end{figure}

Some optimizations are possible if the final graph state is a known quantity before the distribution. We call $G$ the graph representing the graph state to distribute $\ket{G}$. We search to extract from $G$ a star subgraph $S_1$ of maximum size. We distribute the GHZ state associated to $S_1$ using the GHZ state distribution protocol. Then we iterate with $G\setminus S_1$ and so until $\bigcup S_i = G$.

\subsection{Optimality}\label{sec:optimality}
The protocol presented is strictly independent from the network topology and the wanted graph state. The consumption of this general protocol can be compared to the consumption of pathological expensive cases. The most expensive is the case of the line network where we pair each node with its opposite and distribute a Bell pair between them (see Figure \ref{fig:line_case}).
Our protocol gives a consumption of at most $\frac{N(N-1)}{2}$ Bell pairs in $N-1$ time-steps. This upper bound is reached when all the network's nodes are part of the graph state. Both costs are equal up to a factor 2 to the cost of the pathological case (see Table \ref{table}).
 \begin{figure}
     \caption{Depiction of an expensive case. We want to entangle each qubit with the opposite one over a line network.}
     \label{fig:line_case}
     \centering
     \includegraphics[width=.7\columnwidth]{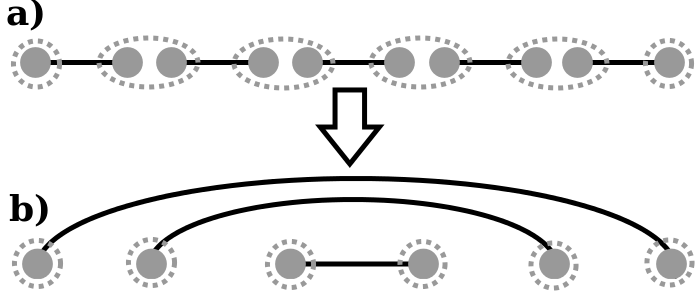}
 \end{figure}
  \begin{table}
    \caption{Creation cost on a network of size $N$} \label{table}
    \begin{tabular}{lccc}
       & & Protocol & Bound\\
    \toprule
      \multirow{2}{*}{$N$-GHZ} & EPR & $N-1$ & $N-1$ \\
      & T & $1$ & $1$\\
        \midrule
    \multirow{4}{*}{Arbitrary Graph State} & \multirow{2}{*}{EPR} & \multirow{2}{*}{$\leq\frac{N(N-1)}{2}$} & \multirow{2}{*}{${\lfloor \frac{N}{2} \rfloor}^2$}\\\\
      & \multirow{2}{*}{T} &  \multirow{2}{*}{$\leq N-1$} & \multirow{2}{*}{$\lfloor \frac{N}{2} \rfloor$}\\\\
      \bottomrule
    \end{tabular}
  \end{table}

\section{Imperfections}\label{sec:imperfections}
Up to this point, we have worked in a lossless and noiseless setting.
Of course, such  idealized setting is far from realistic. 
The full study of multipartite entanglement distribution in the presence of imperfections 
would depend on the details of its physical implementation, and is therefore
beyond the scope of this article. However, several strategies can be discussed to cope with imperfections.

First, the dominant imperfection in a photonic network is expected to be 
losses. In a first approximation, 
they could be modeled by a dynamic network%
, i.e. a network with a graph changing at each time-step, each lost qubit erasing the corresponding edge.
Since our GHZ state distribution protocol is independent of the topology of the graph, as long as all nodes 
are connected, it should be quite robust to losses. Of course, if 
the losses disconnects
the relevant nodes are in subnetworks disconnected by the losses, no GHZ state can
be distributed among them but 
the use of quantum memories and the predistribution of ``partial'' graph states will 
probably allow to bridge the gaps in the next round.
The recent work 
of Khatri et al.\@ \cite{khatri2019practical} on entanglement percolation
lets us hope that such techniques will have modest quantum memory requirements
for large scale quantum networks.

Another important imperfection is the noise itself, it can introduced by the 
needed operations or initially present in the distributed Bell pairs.
A first step to reduce the operation-induced noise is to minimize the number
of operations, a minimization we have not addressed here.
However, since all operations here are Clifford, this kind of optimization is well known 
\cite{browne09computational}.
The noise can also be treated generally with direct purification of the final GHZ or graph states, 
which has been already studied
\cite{kruszynska2006entanglement,aschauer2005multiparticle}.
More generally, many procedures to deal with noise in quantum information, 
including error correction \cite{GottesmanThesis,schlingemann2001quantum}, 
secret sharing \cite{SecretSharing,MarkhamSanders2008}
are based on Clifford operations, and often on graph states \cite{GraphsLong}, 
and quantum fault-tolerant operations are easier within the Clifford group
\cite{Shor1996Fault,Preskill1998Reliable,Gottesman1999Heisenberg}.
They are therefore likely to be compatible with our protocol.

One other idealization in our approach is our neglect of the cost of classical communications
which is already known to be a non-negligible overhead in (classical) network management. 
If the network is static and well known, the Clifford nature of operations allows us to
essentially limit the communication to establishing the path and communicating the
results of the measurements for correction purpose \cite{browne09computational}.
Classical communications in those case should not be too onerous on practice, considering a quantum network will likely have much less traffic than the associated classical one.
However, when losses are taken into account, the network graph itself becomes dynamic
and updating every node about the state of the network will be costly \cite{Pant2019Routing}.
We however hope that quantum secret sharing techniques 
\cite{SecretSharing, MarkhamSanders2008} 
can be used to mitigate this cost.
We also assume all nodes cooperate, and know perfectly the actions of the others. 
In a more realistic setting, imperfect knowledge could lead to congestion issue on the network,
and less altruistic nodes could exploit this. 

\section{Conclusion and perspectives}
\label{sec:conclusion}
We have presented protocols for distributing GHZ and arbitrary graph states which work for networks of any topology.
These protocols are close to optimal in terms of the number of steps $T$ required and the number of Bell pairs consumed in the worst case.

Our model is naturally quite simplified, and there are many possibilities for trade-offs and improvements even within it.
Firstly, we note that the number of steps, $T$,  does not necessarily represent time%
---for example  if nodes are allowed to share $N(N-1)$ Bell pairs, then everything can be done in one physical time step. 
One then has a potential trade-off between how parallel uses of the quantum channel can be, and the use of quantum memory.
Indeed, in terms of memory, one may tweak the steps in our protocol so as to parrallelize as much as possible---with a little thought, one can see one only needs two qubits of memory per Bell pair vertex per node at any one step.
Furthermore, our optimality is for the worst case topology and state---for a fixed graph state and topology one may do much better (e.g. our protocol for the GHZ state). The advantage for our scheme is that it gives a method, and a bound, which works for all states and topology with the same efficiency.

Beyond this, there is much potential for developing more refined  models for quantum networks, where one could for example consider memory, 
local processing  or classical communication costs, as well as their mutual interactions. 
All of these choices can potentially change the optimal strategy, and so must be made carefully. We leave this for future work.
\begin{acknowledgments}
We thank Frederik Hahn, Anna Pappa, Wolfgang Dür and Alexander Priker for
fruitful discussions.
We acknowledge support of the ANR through the ANR-17-CE24-0035 VanQute
project
and from the European Union’s Horizon 2020 research and innovation
programme under grant agreement No 820445 QIA project.
\end{acknowledgments}

\appendix

\end{document}